\documentstyle[twocolumn,eqsecnum,aps]{revtex}

\def\vev#1{\langle #1 \rangle}
\def\half{{\textstyle{1\over2}}}
\def\onebyhbar{{\textstyle{1\over\hbar}}}
\def\ibyhbar{{\textstyle{i\over\hbar}}}
\def\ibytwo{{\textstyle{i\over2}}}
\def\to{\rightarrow}

\def\CC{{\cal C}}
\def\CO{{\cal O}}
\def\CT{{\cal T}}

\def\AMJP#1{{Am. J. Phys.}         {\bf  #1}}
\def\ANP#1{ {Ann. Phys.}           {\bf  #1}}
\def\CMP#1{ {Commun. Math. Phys.}  {\bf  #1}}
\def\IJTP#1{{Int. J. Theor. Phys.} {\bf  #1}}

\def\NCX#1{ {Nuovo Cimento}        {\bf  #1}}
\def\NPB#1{ {Nucl. Phys.}          {\bf B#1}}
\def\PHY#1{ {Physics (NY)}         {\bf  #1}}

\def\PRX#1{ {Phys. Rev.}           {\bf  #1}}

\def\PRL#1{ {Phys. Rev. Lett.}     {\bf  #1}}
\def\PRSA#1{{Proc. R. Soc. London} {\bf A#1}}
\def\RMP#1{ {Rev. Mod. Phys.}      {\bf  #1}}

\begin{document}
\preprint{CTS-IISc/11-98}

\title{On How to Produce Entangled States Violating Bell's Inequalities
in Quantum Theory}
\author{Apoorva Patel\cite{email}}
\address{
CTS and SERC, Indian Institute of Science, Bangalore-560012, India}
\date{\today}
\maketitle

\begin{abstract}
Feynman's path integrals provide a hidden variable description of
quantum mechanics (and quantum field theories). The time evolution
kernel is unitary in Minkowski time, but generically it becomes real
and non-negative in Euclidean time. It follows that the entangled
state correlations, that violate Bell's inequalities in Minkowski time,
obey the inequalities in Euclidean time. This observation emphasises
the link between violation of Bell's inequalities in quantum mechanics
and unitarity of the theory. Search for an evolution kernel that cannot
be conveniently made non-negative leads to effective interactions that
violate time reversal invariance. Interactions giving rise to geometric
phases in the effective description of the theory, such as the anomalous
Wess-Zumino interactions, have this feature. I infer that they must be
present in any set-up that produces entangled states violating Bell's
inequalities.  Such interactions would be a crucial ingredient in a
quantum computer.
\end{abstract}

\narrowtext

\section{Feynman's path integral\protect\\ as a hidden variable theory}

It is common, and often more convenient, to study quantum mechanics
using the Schr\"odinger/Dirac equations and the Heisenberg operator
algebra. Feynman's path integral formulation of quantum mechanics
\cite{Feyn65},
nonetheless, is an alternative approach from which all the known
results of quantum mechanics can be derived. In fact, quantum field
theories are studied, more often than not, using the path integral
formulation. For simplicity, let us consider quantum mechanics in
one space dimension. The total amplitude (or wavefunction) for the
system to be in the state $\psi(x_f,T)$ at time $t=T$, given an
initial state $\psi(x_i,0)$, is defined in terms of the transition
kernel $K(x_f,T;x_i,0)$:
\begin{equation}
\psi(x_f,T) ~=~ \int_{-\infty}^\infty K(x_f,T;x_i,0)
              ~ \psi(x_i,0) ~dx_i ~~,
\end{equation}
\begin{equation}
K(x_f,T;x_i,0) ~=~ \int_0^T [Dx(t)] ~\exp [\ibyhbar S(x(t))] ~~.
\end{equation}
Here the action for a particular trajectory is the time integral of
the corresponding Lagrangian,
\begin{equation}
S(x(t)) ~=~ \int_0^T L(x(t))~dt ~~.
\end{equation}
The integration measure $[Dx(t)]$ can be defined precisely by
discretising the time interval: 
\begin{equation}
[Dx(t)] ~\propto~ \lim_{N\to\infty} \int_{-\infty}^\infty
         \prod_{j=1}^{N-1} dx_j ~~,~~ x_j \equiv x(t=jT/N) ~~.
\end{equation}
Transition matrix elements and other expectation values are defined
with the straightforward prescription:
\begin{eqnarray}
\vev{\CO(x(t))} &&~\equiv~ \vev{\psi(x_f,T) | \CO(x,t) | \psi(x_i,0}
		\nonumber\\
                &&~=~ \int [Dx(t)] ~\rho_M (x(t))~ \CO ~~,
\end{eqnarray}
\begin{equation}
\rho_M (x(t)) ~=~ {\exp [\ibyhbar S(x(t))] \over
                  \int [Dx(t)] ~\exp [\ibyhbar S(x(t))]} ~~.
\end{equation}

These definitions provide a hidden variable description of quantum mechanics.
Indeed, $x_j~(j=1,...,N-1)$ are the hidden variables which are integrated
over \cite{footA}.
There is no need to worry about ordering of various factors, because there
are no non-commuting operators in this Lagrangian description, only complex
numbers \cite{footB}.
The functional integration measure $[Dx(t)]$ represents a sum over all paths
connecting the fixed initial and final states; such a sum over paths is
inherently a non-local object \cite{footC}.
$\exp[\ibyhbar S(x(t))]$ is the statistical weight of the path $x(t)$
which depends on the interaction amongst the particles in the system.
It is well-known that the typical paths contributing to the functional
integral are highly irregular and non-differentiable. The individual
paths characterised by $\{x_j\}$ follow local history/dynamics \cite{footD},
but they do not obey constraints of causality. The virtual intermediate
states, for example, can be ``off-shell'' with no relation between their
energies and momenta, and they can propagate at a speed faster than that
of light. They can even propagate backwards in time which is interpreted
as pair creation and annihilation in relativistic field theories.
The total amplitude (i.e. the sum over all paths), however, obeys all
the constraints of causality and conservation laws.

Note that $\{x_j\}$ corresponding to one particle are totally independent
of those for another. For a system of non-interacting particles, the path
integral completely factorises, e.g.
$[Dx(t)] = [Dx^{(1)}(t)]~[Dx^{(2)}(t)]$ and
$S(x(t)) = S(x^{(1)}(t)) + S(x^{(2)}(t))$.
In such a case, all the correlations/entanglements amongst the particles
are built into the specification of the initial state coordinates,
$x_i^{(n)}$. The subsequent evolution of the system and measurement of
single particle properties clearly separate into independent components
corresponding to each particle. Such a separation, often dubbed the
``Einstein locality'' property of the hidden variables, is an important
ingredient in the proof of Bell's inequalities \cite{Bell64}. It ensures
that what is measured for one particle is not all influenced by either
the property being measured or the choice of the measurement apparatus
for the other particle. It has to be stressed that even though the path
integrals are non-local while describing single particle evolution,
they comply with Einstein locality while describing inter-particle
correlations \cite{footE}.

The measurement process here corresponds to restricting the sum over all
possible paths to only those paths which are consistent with the measured
observable having a specific value in the final state. This is obviously
a contextual process, once all the hidden variables are integrated out;
if a second measurement were to be carried out on the system, the sum
over all paths would be restricted to paths which are consistent with the
results of both the first and the second measurements---the paths which
would be consistent with the second measurement but not with the first
have been discarded by the act of the first measurement. Thus the measured
value of an observable may depend on the results of other measurements
carried out prior to it on the same system, when these other measurements
correspond to commuting but correlated observables \cite{footF}.
Note that the order in which the restriction of paths is carried out is
immaterial, so there is no ambiguity regarding the final state of the
system when measurements are carried out at space-like separations. 

All put together, path integrals describe a contextual and non-local
hidden variable theory; as a matter of fact, a non-zero value of $\hbar$
and Bell's theorems ensure that any hidden variable theory describing
quantum mechanics has to have such peculiarities \cite{Mermin93}.
It is similar in many aspects to the de Broglie-Bohm theory \cite{Bohm52},
but with the clear advantage that it is describable in a simpler language
and that it is much more amenable to detailed calculations \cite{footG}.
The crucial feature in this description is that the integration weight,
$\rho_M (x(t))$, is a complex number in general. Therefore, although it
is bounded, it cannot be interpreted as a probability density. It is
this fact which allows the path integral description to bypass Bell's
inequalities and give a true definition of quantum mechanics \cite{footH}.

\section{Wigner function and phase space formulation}

It is often considered desirable (although it is not at all necessary for
the proof of Bell's inequalities) that a hidden variable description of
quantum mechanics would provide simultaneous reality to non-commuting
physical observables. For example, simultaneous statistical weight can be
given to positions and momenta by formulating the theory in the phase space.
Feynman actually showed how one can reconcile the EPR paradigm \cite{EPR35}
with quantum mechanics using hidden variables, provided that probabilities
are allowed to become negative \cite{Feyn82}. He used the well-known Wigner
function \cite{Wigner}, which is a particular realisation of the density
matrix distribution in the phase space:
\begin{equation}
W(x,p) ~=~ \int ~dy~ \psi^* (x+\half y) ~\exp (\ibyhbar py)
                   ~ \psi   (x-\half y) ~~.
\end{equation}
Wigner functions are real \cite{footI} and obey all the usual manipulations
of probability theory, except that they are not always non-negative
everywhere in the phase space. Since the expectation value for any
physical observable is just
\begin{equation}
\vev{\CO} ~=~ \int dx~dp ~ W(x,p) ~ O(x,p) ~~,
\end{equation}
where $O(x,p)$ is the (Hermitian) operator weight corresponding to the
observable $\CO$, it is necessary that for situations violating Bell's
inequalities the Wigner function be negative somewhere in the phase space.

Path integrals can also be formulated in the phase space. The transition
kernel is expressed as
\begin{equation}
K(x_f,T;x_i,0) ~=~ \int_0^T [Dx][Dp] ~\exp [\ibyhbar S(x,p)] ~~,
\end{equation}
with the action rewritten in terms of the Hamiltonian as
\begin{equation}
S(x,p) ~=~ \int_0^T [p {dx \over dt} - H(x,p)]~dt ~~.
\end{equation}
Though this is the more general formulation, it has become customary (and
convenient) to completely integrate out either the $\{x_j\}$ or the $\{p_j\}$
variables and work with the Lagrangian description. Generic operators
$O(x,p)$ need to be converted to either the $x$- or the $p$-language, using
Fourier transforms wherever necessary. There is no loss of predictive power,
but some care is needed in time-ordering products of non-commuting operators.
For instance, the commutation relation, $[\hat{x},\hat{p}]\ne0$, is realised
as \cite{Feyn65}:
\begin{equation}
\lim_{N\to\infty} \vev{x_j~m{x_j-x_{j-1} \over T/N}
		  ~-~ m{x_{j+1}-x_j \over T/N}~x_j} ~\ne~ 0 ~~.
\end{equation}

Apart from having a fewer number of variables to deal with, a striking
advantage of the Lagrangian formulation is that the initial states of
variables can be specified freely, subject to only the normalisation
constraint. On the other hand, initial state phase space distributions
cannot be arbitrary for physical situations; they have to satisfy
restrictions following for example from the uncertainty principle,
$\vev{\Delta x}\vev{\Delta p} \ge {\hbar\over2}$.

\section{Role of unitarity}

Now let us apply the familiar trick of rotating to Euclidean (imaginary)
time, $\tau=it$. This Wick rotation converts the quantum theory to the
language of statistical mechanics. The integration weight,
\begin{equation}
\rho_E (x(\tau)) ~=~ {\exp [-\onebyhbar S(x(\tau))] \over
                     \int [Dx(\tau)] ~\exp [-\onebyhbar S(x(\tau))]} ~~,
\end{equation}
is now real and non-negative, and can be interpreted as a probability
density. This formal property has been exploited before, for vector-like
gauge theories at zero chemical potential \cite{footJ},
to derive rigorous inequalities among correlation functions and particle
masses \cite{Weingarten,VafaWitten}.
The point I want to emphasise in this article is that a non-negative
integration weight must obey Bell's inequalities \cite{footK}.

Suppose that we can unambiguously predict the Minkowski time results
from Euclidean time ones. Then we have a hidden variable prescription
for describing quantum mechanics. This cannot be correct. We have to
identify ways out of this situation, and they will lead us to the origin
of the violations of Bell's inequalities. There are two options:
(1) Something is lost in the analytic continuation from Minkowski to
Euclidean time, which prevents complete reconstruction of Minkowski
time results from Euclidean time ones. (2) The integration weight
obtained by analytic continuation to Euclidean time is not non-negative,
may be even complex. I explore this possibility in section V.

Let us first note that the above mentioned analytic continuation
is routinely employed in quantum field theories in dealing with
divergent loop integrals and renormalisation. As a matter of fact,
there are strong theorems governing such an analytic continuation,
e.g. the Wightman axioms \cite{Streater}
and the Osterwalder-Schrader positivity of the transfer matrix
\cite{Osterwalder}.
In particular, in the complex energy plane, these theorems rely on a
sufficiently fast decrease of the amplitudes at infinity and on there
being no singularities in the region covered by the rotation \cite{footL}.
The integration contours can then be freely rotated without affecting
the value of the integrals.

Let us also note that statistical mechanics has many features in
common with quantum theory. The density matrix description allows
probabilistic interpretation of superposed and mixed states. The
presence of the heatbath \cite{footM}
gives rise to vacuum fluctuations and unconstrained behaviour of the
virtual states. Non-zero commutators (Poisson brackets) leading to
the uncertainty principle, e.g. $[\hat{x},\hat{p}]\ne0$,
exist in the Hamiltonian description of the theory. Non-zero tunnelling
amplitudes exist in the Euclidean time theory, and so do ``grotesque''
states with infinitesimal probabilities \cite{footN}.
These shared features cannot be responsible for the violation of
Bell's inequalities in quantum mechanics.

What the Euclidean time theory lacks is the key concept of unitarity.
The Minkowski time transition kernel, $K_M (x_f,T;x_i,0)$, corresponds
to a unitary transformation---the familiar $S-$matrix \cite{footO}.
It is easy to see that a unitary matrix with only real and positive
matrix elements has to be the identity matrix (or its row-wise
permutation corresponding to a shuffling of the states). The Euclidean
time transition kernel is less restricted---it does not preserve the
norm, though it maintains orthogonality of the states---and can be
represented by a diagonal non-negative definite matrix. The loss of
normalisation is not critical, since it can be taken care of
following the LSZ prescription \cite{LSZ}. The information about
the relative phases of the states, however, is lost.
In fact this is the crucial quantum mechanical feature which makes the
requirements of unitarity and a real non-negative integration weight
mutually incompatible. One can pick situations where the information
contained in the relative phases cannot be made arbitrary small,
because there are constraints on the complex amplitudes following,
for example, from analyticity and dispersion relations \cite{footP}.
In such cases, a statistical mechanics description cannot provide
an arbitrarily close approximation to quantum mechanics. Correlations
violating Bell's inequalities can be present in such situations.

Let us look at the problem again from a slightly different angle.
The Euclidean time correlation functions defined along the imaginary
time axis are real, and a mere analytic continuation of them
cannot produce non-trivial complex phase shift factors that are an
an essential part of Minkowski time scattering amplitudes \cite{footQ}.
{\it A priori} one does not know whether in analytical continuation
of the Euclidean time results the Euclidean time axis should be rotated
by $+90^\circ$ or $-90^\circ$ to reach the Minkowski time axis.
This ambiguity automatically gets resolved for internal loop
variables of Feynman diagrams, just due to the necessity of not
crossing any singularities while rotating the integration contours.
For the variables corresponding to the external legs (incoming and
outgoing states), however, the choice must be enforced as an additional
condition; it amounts to the difference between choosing advanced or
retarded propagators. Note that given the Euclidean time results,
one can construct either the advanced or the retarded (or a linear
combination thereof) amplitudes. The criterion of retarded amplitudes
only, though physically motivated by principle of causality, is absent
in the Euclidean time theory and has to be enforced as an additional input.

More explicitly, the principle of causality is embedded in the use
of $i\epsilon-$prescription for retarded propagators. For example,
the naive analytic continuation of the Euclidean time scalar boson
propagator, $(k^2+m^2)^{-1}$, to $(k^2-m^2)^{-1}$ is incorrect.
The correct physical prescription is:
\begin{eqnarray}
{1 \over k^2+m^2} ~\to&&~ \lim_{\epsilon\to0}
                          {1 \over k^2-m^2+i\epsilon} \nonumber\\
                    ~=&&~ {\rm P} \big( {1 \over k^2-m^2} \big)
                          - i\pi\delta(k^2-m^2) ~~.
\end{eqnarray}
The non-unitary Euclidean time theory fixes the off-shell amplitude
(the principle value part) completely, but it is necessary to add
an on-shell contribution (exemplified by the $\delta-$function at
the pole) to comply with unitarity. This subtlety is unimportant
in many cases: location of poles and branch cuts, matrix elements
with external legs amputated according to the LSZ prescription,
scattering amplitudes at threshold which are real etc. can be
determined without recourse to the $i\epsilon-$prescription. Indeed
Monte Carlo simulations of quantum field theories on the lattice
calculate all such quantities in the Euclidean time framework.
On the other hand, the features that can differentiate between advanced
and retarded amplitudes (e.g. $S-$matrix phase shifts, discontinuities
across branch cuts etc.) are not directly accessible in the Euclidean
time framework.

\section{Analysis of familiar examples}

\subsection{Original EPR correlations}

The conventional set up illustrating Bell's inequalities is the case
where there is just free propagation of particles after initial
creation of the state. Let us first look at the original EPR example
\cite{EPR35}---a system of two identical non-relativistic particles
freely propagating in one space dimension, where the initial state
of the two particles is perfectly correlated in space (hence
anti-correlated in momentum) with the constraint
\begin{equation}
\delta (x_i-y_i) ~=~ \int {dp \over 2\pi\hbar}
                   ~ \exp \big( \ibyhbar p(x_i-y_i) \big) ~~.
\end{equation}
For a single free particle of mass $m$, the propagation kernel for
Minkowski time $T$ is
\begin{equation}
K_M (x_f,T;x_i,0) ~=~ \sqrt{m \over {2\pi i\hbar T}}
  ~ \exp \big( i{m(x_f-x_i)^2 \over {2\hbar T}} \big) ~~,
\end{equation}
while for Euclidean time $T$ it is
\begin{equation}
K_E (x_f,T;x_i,0) ~=~ \sqrt{m \over {2\pi\hbar T}}
  ~ \exp \big( -{m(x_f-x_i)^2 \over {2\hbar T}} \big) ~~.
\end{equation}
Thus the final state wavefunction for the two particle system is
\begin{equation}
\psi_M (x_f,y_f,T) ~=~ \int {dp \over 2\pi\hbar}
                       \exp \big( \ibyhbar p(x_f-y_f) \big)
                       \exp \big( {ip^2 T \over \hbar m} \big)
\end{equation}
in Minkwoski time, and
\begin{equation}
\psi_E (x_f,y_f,T) ~=~ \int {dp \over 2\pi\hbar}
                       \exp \big( \ibyhbar p(x_f-y_f) \big)
                       \exp \big( -{p^2 T \over \hbar m} \big)
\end{equation}
in Euclidean time. In both cases the theory maintains its
contextual character; the structure of the kernel ensures
that if one particle is detected with momentum $p_f$,
the other is bound to be found with momentum $-p_f$.

The relative probability of observing various values of momentum
is different in the two cases, however, which is just due to the
difference in normalisation of states. This difference is not small
as can be seen by appealing to the uncertainty relation: In order
to detect the two particles distinctly, their separation has to be
much larger than their de Broglie wavelengths,
\begin{equation}
|x_f-y_f| ~\sim~ p_f T/m ~\gg~ \hbar/p_f ~~,
\end{equation}
implying that $p^2 T \gg \hbar m$ in the exponent. In this particular
example, the exponent can be removed by just following the LSZ
prescription, and there is no conflict with any inequality.

The reason behind no conflict with any inequality in the above
example is that the $\delta-$function correlation is non-negative.
A more general case \cite{Bell86}
will have initial state correlations such that the Wigner function
becomes negative somewhere in the phase space \cite{footR}.
The density matrix evolves linearly in time according to:
\begin{eqnarray}
{d W_M \over dt}    &&~=~ -\ibyhbar   [H,W_M] ~~\Longrightarrow~~ \nonumber\\
  && W_M(t)    ~=~ \exp(-\ibyhbar   Ht)    W_M(0)
                   \exp( \ibyhbar   Ht) ~~, \nonumber\\
{d W_E \over d\tau} &&~=~ -\onebyhbar [H,W_E] ~~\Longrightarrow~~ \nonumber\\
  && W_E(\tau) ~=~ \exp(-\onebyhbar H\tau) W_E(0)
                   \exp( \onebyhbar H\tau) ~~,
\end{eqnarray}
where $H$ is the Hamiltonian for the system.
In case of a non-negative time evolution kernel, interference effects
can only annihilate the negative density matrix regions in the phase
space; a limit is reached when the Wigner function becomes non-negative
everywhere and thereafter no regions of negative Wigner function can
be regenerated \cite{footS}.
Explicitly
\begin{eqnarray}
f(\tau) &&~=~ \int dx~dp ~ |W(x,p;\tau)| ~/~ \int dx~dp ~ W(x,p;\tau) ~~,
              \nonumber\\
f(\tau) &&\ge 1 ~~,~~ df(\tau)/d\tau \le 0 ~~.
\end{eqnarray}
No such reduction of negative density matrix regions is expected in case
of a unitary time evolution; in fact the Minkowski time evolution of the
density matrix obeys the continuity equation \cite{Wigner}:
\begin{equation}
{\partial W(x,p) \over \partial t}
  + {p \over m} {\partial W(x,p) \over \partial x} ~=~ 0 ~~.
\end{equation}
Thus, after factoring out the normalisation of states, the residual
correlations which typify Bell's inequalities, become substantially
different in the Euclidean time case from the corresponding Minkowski
time values. It can be argued that a sense of probability description
can be retained in Euclidean time, if one replaces
$\psi(t)\rightarrow\psi(\tau)$ and $\psi^*(t)\rightarrow\psi^*(-\tau)$
\cite{Shamir}. But the resultant expression is so non-local in
Euclidean time that it is not possible to assign physical meaning to it.


To summarise, the Euclidean time evolution cannot dynamically create
correlations violating Bell's inequalities if such correlations are
absent in the initial state. Moreover, it wipes out such correlations
even when they are inserted by hand in the initial state.

\subsection{Bohm-Aharanov correlations}

Violation of Bell's inequalities in quantum mechanics is often
demonstrated using two non-relativistic spin-$\half$ particles
in a singlet state \cite{BohmAharanov}.
It is well-known that the Wigner function for this state has
negative elements \cite{Feyn82,footT}.
A path integral representation to $SU(2)$-spin dynamics can be given
in terms of coherent states \cite{FradkinPerelomov}.
These coherent states are represented by unit vectors $\vec{n}$ from the
origin to points on a two-dimensional unit sphere $S_2$ \cite{footU}.
The conventional spin-$\half$ eigenstates are represented by the unit
vectors pointing to the north and south poles:
\begin{equation}
|\uparrow\rangle = |\vec{n}_0\rangle ~,~
|\downarrow\rangle = |-\vec{n}_0\rangle ~.
\end{equation}
A general coherent state is obtained by rotating the reference vector
$\vec{n}_0$ to $\vec{n}$,
\begin{equation}
|\vec{n}\rangle ~=~ \exp(\ibytwo\theta\vec{a}\cdot\vec{\sigma})
|\uparrow\rangle ~,
\end{equation}
where $\vec{a}$ is the unit vector in the direction $\vec{n}_0\times\vec{n}$,
and $\sigma_i$ are the usual Pauli matrices. The set of coherent states is
overcomplete, with a uniform and positive integration measure over $S_2$.
One can revert back to the conventional basis by integrating over the polar
angles $\theta,\phi$ parametrising $S_2$.

In this coherent state picture, the spin-dynamics is described by the
time evolution $|\vec{n}(t)\rangle$. For every spin, the Minkowski time
action contains a topological Wess-Zumino term:
\begin{eqnarray}
S_M(\vec{n}) ~=~&& \half S_{WZ}(\vec{n}) ~+~ {i\delta t\over8}
		   \int_0^T dt~(\partial_t \vec{n}(t))^2\nonumber\\
	     ~+~&& S_{interaction} ~~.
\end{eqnarray}
For a spin in a magnetic field,
$S_{interaction} \propto \half\int_0^T dt~\vec{n}\cdot\vec{B}$.
The Wess-Zumino term is best expressed by embedding the system into one
higher dimension,
\begin{equation}
S_{WZ}(\vec{n}) ~=~ \int_0^1 ds \int_0^T dt ~
[\vec{n}(t,s)\cdot\partial_t\vec{n}(t,s)\times\partial_s\vec{n}(t,s)] ~~,
\end{equation}
where $\vec{n}(t,0)\equiv\vec{n}(t)$, $\vec{n}(t,1)\equiv\vec{n}_0$.
The evolution kernel for a single free spin is given by the propagator,
\begin{eqnarray}
&& K_M (\vec{n}_f,T;\vec{n}_i,0) ~=~ \vev{\vec{n}_f|\vec{n}_i}\nonumber\\
&& ~=~ \exp(\ibytwo \Phi(\vec{n}_i,\vec{n}_f,\vec{n}_0))
     ~ \sqrt{\half(1+\vec{n}_f\cdot\vec{n}_i)} ~~,
\end{eqnarray}
where $\Phi(\vec{n}_i,\vec{n}_f,\vec{n}_0)$ is the area of the geodesic
spherical triangle formed by $\vec{n}_i,\vec{n}_f,\vec{n}_0$. The first
factor in the kernel is a unitary phase coming from the Wess-Zumino term,
while the second factor is non-negative.

A peculiarity of the Wess-Zumino term is that it contains an odd number
of time derivatives (the action in the previous subsection had only
even number of time derivatives). As a result, its contribution to the
exponent of the path integral weight is imaginary in both Minkowski and
Euclidean times. After Wick rotation, the Euclidean time kernel is still
a complex number and cannot be interpreted as a probability density.
As a matter of fact, for a single free spin, the Minkowski and Euclidean
time evolution kernels coincide,
$K_M (\vec{n}_f,T;\vec{n}_i,0) ~=~ K_E (\vec{n}_f,T;\vec{n}_i,0) ~$.

Going back from the coherent state basis to the conventional one, it is
easily seen that a free spin state remains unchanged under time evolution.
Thus the entangled singlet state violating Bell's inequalities,
$|\uparrow\downarrow\rangle - |\downarrow\uparrow\rangle$, remains the
same under both Minkowski and Euclidean time evolutions.

At a deeper level, we can enquire how the unusual Wess-Zumino term came
about. Spin is a property of the Poincar\`e group transformations
describing free particles. Expressing the spin operator as the
Pauli-Lubansk\`{\i} $4-$vector,
$m s^\mu = \half\epsilon^{\mu\nu\sigma\tau} p_\nu J_{\sigma\tau}$,
(here $J_{\sigma\tau}$ are the Lorentz generators), we notice that it
has to transform as an axial vector under Wick rotation. A complete
implementation of Minkowski to Euclidean time rotation described above
would transform
$\vec{s} \rightarrow -i\vec{s}$ (or $\vec{n} \rightarrow -i\vec{n}$),
as appropriate for an angular momentum. Although such a transformation
would apparently make the Euclidean time kernel non-negative, it totally
destroys the underlying unitary structure of the rotation group $SU(2)$
\cite{footV}.
(The same transformation is also required to make the interaction term
of the spin with a magnetic field real in Euclidean time; under Wick
rotation the magnetic field, generated by currents, transforms as
$\vec{B}\rightarrow -i\vec{B}$.)
It is not at all clear whether such an analytic continuation has any
mathematical meaning, or whether such a positive counterpart to $SU(2)$
has any physical interpretation.
%
%

Historically, Wess-Zumino type of interactions have been labeled
anomalous. This is largely because they violate apparently good
symmetries; resolution of the puzzle requires a proper account of
the internal properties of the variables. It is worthwhile to
recollect that anomalies express global properties of the system
and cannot be eliminated by local transformations \cite{anomalies},
a property also shared by entangled states. The classic example
is that of a neutral pion decaying into two entangled photons in a
singlet state. A Wess-Zumino term describes this decay \cite{Witten}.
The properties of this term under Wick rotation are linked with the
facts that the pion field is a pseudoscalar and parametrises a unitary
group manifold.

\section{Origin of entanglement}

We are now ready to characterise what type of interactions would give
rise to the entanglements that violate Bell's inequalities. Normally
Bell's inequalities are not studied from this view point. They are
instead specified by certain correlations in the initial state of a
non-interacting system. In the real world, say in any $S-$matrix
set-up where there are no correlations whatsoever in the initial state,
we have to analyse how the interactions dynamically generate the
correlations appearing in the final state. The question becomes much
more important, when one takes the view that correlations are all there
is to quantum theory \cite{Mermin98}.

An initial state without correlations can easily be described by a
Wigner function non-negative everywhere in the phase space. In such a
case, we have seen above that a non-negative evolution kernel cannot
create correlations violating Bell's inequalities. It is easy to see
that with a time reversal invariant Lagrangian, the Wick rotation
converts a Minkowski time kernel to a non-negative Euclidean time kernel:
\begin{eqnarray}
iS_M &&= i\int dt~L_M(t)       ~,~ L_M(t) = L_M(-t)\nonumber\\
~\Longrightarrow~
-S_E &&= -\int d\tau~L_E(\tau) ~,~ L_E(\tau) = L_E(\tau^*) ~~.
\end{eqnarray}
Needless to say, complexification of time provides a continuous route
to reach a time-reversed configuration; Wick rotation is just proceeding
half way along this route.

We note that though violation of time reversal symmetry, $\CT$, is present
in the standard model of interactions of fundamental particles \cite{footW},
it is tiny and inconsequential in problems involving violations of Bell's
inequalities. There also exist Boltzmann's famous $H-$theorem describing
monotonic increase of entropy, and dissipative terms that come about from
interaction of the system with its environment, but such irreversibility
is of no concern for the problem at hand (unitary evolution in quantum
theory is reversible). We thus restrict ourselves to situations where the
fundamental underlying theory is time reversal invariant. It is important
to realise that this assumption does not forbid localised arrangements
that do not respect time reversal symmetry. For example, a magnetic field
is not time reversal invariant, although the underlying QED is, and we can
very well produce magnetic fields in our laboratories.

Several examples of interactions that respect unitarity of the evolution
kernel, and yet violate $\CT$, are known. Generically they can be described
in terms of geometric phases (also called Berry's adiabatic phases)
\cite{Berry,ShapereWilczek}:
\begin{equation}
iS_{Geom}(T) ~=~ i\int_0^T dt \vec{A}(\vec{R})\cdot{d\vec{R}\over dt}
             ~=~ i\int_0^T \vec{A}(\vec{R})\cdot d\vec{R} ~~.
\end{equation}
Here $\vec{R}$ labels the states in the quantum Hilbert space of the system,
and $\vec{A}$ is an effective gauge potential in this space. Different
physical situations are described by different group theoretical structure
and holonomy of $\vec{A}$. Examples are \cite{ShapereWilczek}:
Pancharatnam's and Guoy's phases in optics, 
correction phase to Born-Oppenheimer approximation in atomic/molecular
physics, Aharanov-Bohm phase in electrodynamics, 
and Wess-Zumino terms describing anomalous interactions. All these phases
encode global properties of the system, a feature that is mandatory for
describing entangled states.

The situation is best described in the framework of effective theories;
all the degrees of freedom of the fundamental theory that are not observed
and that are of no direct interest are summed over (say using the path
integral formalism). The remaining observed degrees of freedom may be
composite in terms of the fundamental ones. The underlying fundamental
theory dictates the nature of these effective degrees of freedom and the
effective interactions they have amongst themselves. Effective interactions
describable by a geometric phase become possible, only when the effective
degrees of freedom possess a unitary symmetry as well as nontrivial $\CT$
transformation property in the effective theory. Within the domain of the
effective theory, the internal properties of the effective degrees of
freedom are not available for external manipulations. We can therefore
redefine $\CT$ as acting only on the effective interactions, while leaving
the internal properties of the effective degrees of freedom untouched.
The effective interactions can then create correlations violating
Bell's inequalities amongst the effective degrees of freedom. Wess-Zumino
terms are an important example of such interactions. Given the existence
of these anomalous interactions in the examples above, we can even say that
$\CT-$invariance of Poincar\`e group transformations for free particles
requires the spin to be an axial vector, and $\CT-$invariance of QCD
requires the pion to be a pseudoscalar.

Correlations violating Bell's inequalities can also be in the internal
symmetry space of a system, e.g. isospin and colour correlations amongst
quarks making up mesons and baryons. In such cases, the charge conjugation
symmetry of the appropriate internal symmetry group, $\CC$, plays the same
role that the time reversal symmetry, $\CT$, did in the above analysis.
(In essence, the effective theory description above combined a unitary
symmetry with the internalised time reversal property, in such a way that
the combination transforms the same way under $\CT$ as it would under
$\CC$ for the unitary symmetry.) The crucial connection with unitarity
is maintained because the symmetry groups in quantum theory are unitary
or subgroups of unitary groups \cite{footX}.
These internal space correlations, unfortunately, are not directly
accessible for creation/verification in an experimental set-up.

Finally we observe that the scattering phase shifts discussed in section III
change sign under time reversal; a feature of the amplitude that can
distinguish between advanced and retarded behaviour must be odd under $\CT$.

\section{Summary and outlook}

The purpose of this article is to highlight some mathematical features
of quantum mechanics crucial to creation of entangled states that
violate Bell's inequalities. The deeper philosophical issues have
been kept in the background, and explanation of technicalities has
been relegated to footnotes. It is impertinent to ask why nature opted
for unitarity or the Minkowski metric. What I have aimed to emphasise
is that it is the ubiquitous appearance of ``$i$'' together with a
non-zero value of ``$\hbar$'', so characteristic of quantum physics,
that sets it apart from classical physics: ``$i$'' is responsible for
correlations violating Bell's inequalities, while ``$\hbar\ne0$'' is
responsible for non-locality. A lesson to be learned is that care must
be exercised in converting results of Euclidean time field theory to
physical amplitudes; the restrictions following from unitarity have
to be kept in mind.

From the practical viewpoint, effective interactions that do not obey
the discrete $\CT$ symmetry would be indispensable in a quantum computer.
After all, the end result of a quantum computation is nothing but a
correlation between the input and the output. In brief, a quantum computer
operates on qubits ($2-$state quantum systems) with unitary operations, and
achieves speed-up over its classical counterpart by clever implementation
of superposition/phase-rotation and entanglement/interference.
While superposition can be easily achieved by preparing the quantum
state in one basis and afterwards using/observing it in another,
entanglement of initially uncorrelated quantum bits would require the
type of ``anomalous'' interactions discussed above. A typical component
of a quantum computer is the $2-$qubit controlled-not operation (quantum
generalisation of the Boolean exclusive-OR), which can convert
superposition into entanglement:
\begin{equation}
(|\uparrow\rangle+|\downarrow\rangle)|\downarrow\rangle
~\mathrel{\mathop{\longrightarrow}^{\rm controlled}_{\rm not}}~
|\uparrow\rangle|\uparrow\rangle+|\downarrow\rangle|\downarrow\rangle ~~. 
\end{equation}
It has been implemented using scattering phase shifts \cite{Expt1},
and effective spin-$\half$ state representations \cite{Expt2,Expt3}.

\acknowledgments

This article is an expansion of an earlier unpublished work \cite{bellineq93}.
I am grateful to N. Mukunda for many discussions. I thank John Preskill and
Joseph Samuel for their comments on the preliminary version of this article.
This work was supported in part by the Rajiv Gandhi Institute of Contemporary
Studies in cooperation with the Jawaharlal Nehru Centre for Advanced Scientific
Research, Bangalore.


\begin{references}
\bibitem[*]{email} E-mail: adpatel@cts.iisc.ernet.in
\bibitem{Feyn65} R. Feynman, \RMP{20}, 367 (1948); \hfil\break
              R. Feynman and A. Hibbs, {\it Quantum Mechanics
              and Path Integrals} (McGraw-Hill, New York, 1965).
\bibitem{footA} Hidden variables are commonly interpreted to be as yet
              undiscovered internal degrees of freedom. But this need
	      not be so. As far as Bell's theorems are concerned,
	      hidden variables can be ``any'' set of variables, internal
	      or external, that are summed over and cannot be observed.
\bibitem{footB} Fermions can be dealt with along similar lines using
	      Grassmann variables, but I do not consider them here.
\bibitem{footC} The integration measure can be made local by suppressing
	      the quantum fluctuations in the formal limit, $\hbar\to0$.
	      When referring to a single object, my use of the adjective
	      ``local'' or ``non-local'' means whether at a fixed instant
	      of time it is point-like or spread over a region.
\bibitem{footD} This is true in quantum mechanics only for non-interacting
              particles, since long range inter-particle interaction
	      potentials would produce instantaneous action-at-a-distance.
	      The more general framework of quantum field theories can
	      produce instantaneous action-at-a-distance from local
	      Lagrangian densities by exchange of virtual particles.
\bibitem{Bell64} J. Bell, \PHY{1}, 195 (1964); \hfil\break
              J. Bell, {\it Speakable and Unspeakable in Quantum
              Mechanics} (Cambridge University Press, Cambridge, 1987).
\bibitem{footE} The use of the same adjective ``local'' to describe both
	      single particle particle evolution and inter-particle
	      correlations is unfortunate, but I have no option short of
	      inventing a new phrase.
\bibitem{footF} This way of restricting the sum over paths is similar to
	      Bayes' rule for conditional probabilities.
\bibitem{Mermin93} A recent review on the implications of Bell's theorems
              can be found in: N. D. Mermin, \RMP{65}, 803 (1993).
	      ``Physical locality'' is obeyed in quantum mechanics for all
	      observable quantities (i.e. expectation values and correlations),
	      and no real signal can propagate faster than the speed of light.
	      ``Quantum non-locality'' refers to the behaviour of presumed
	      unobservable objects used in the mathematical formulation of
	      quantum mechanics (e.g. states, wavefunctions, path trajectories
	      etc.). Such hidden variables crop up when actual results of
	      experiments are compared with hypothetical results of
	      experiments that could have been performed but were not
	      performed, and Bell's theorems demand that these hidden
	      variables be non-local. See also Ref.~\onlinecite{Mermin98}.
\bibitem{Bohm52} D. Bohm, \PRX{85}, 166 (1952); 180 (1952).
\bibitem{footG} In particular, we know how to generalise from
	      non-relativistic to relativistic case, and from quantum
	      mechanics to quantum field theory.
\bibitem{footH} For example, there can be cancellation between different
              paths giving rise to interference effects.
\bibitem{EPR35} A. Einstein, B. Podolsky and N. Rosen,
              \PRX{47}, 777 (1935); \hfil\break
              A. Einstein, in {\it Albert Einstein, Philosopher Scientist},
	      edited by P. Schilp, Library of Living Philosophers
              (Evanston, Illinois, 1949).
\bibitem{Feyn82} R. Feynman, \IJTP{21}, 467 (1982); \hfil\break
              R. Feynman, ``{\it Negative probability}'' in
              {\it Quantum Implications: Essays in honour of
              David Bohm}, edited by B. Hiley and F. David Peat,
              (Routledge \& Kegan Paul, London, 1987), p. 235.
\bibitem{Wigner} E. Wigner, \PRX{40}, 749 (1932).
\bibitem{footI} The density matrix is a Hermitian operator, and can always
	      be made real by a suitable choice of basis. Thereafter, 
	      orthogonal transformations are sufficient to describe its
	      evolution.
\bibitem{footJ} QED and QCD belong to this class, in particular the example
	      of para-positronium or neutral pion decaying into two correlated
	      photons.
\bibitem{Weingarten} D. Weingarten, \PRL{51}, 1830 (1983).
\bibitem{VafaWitten} C. Vafa and E. Witten, \NPB{234}, 173 (1984); \hfil\break
              E. Witten, \PRL{51}, 2351 (1983).
\bibitem{footK} I use the term ``Bell's inequalities'' in this article
	      to generically denote the type of inequalities proved by Bell
	      as well as by others.
\bibitem{Streater} R. Streater and A. Wightman, {\it PCT, spin and
              statistics, and all that} (Benjamin/Cummings, New York, 1964).
\bibitem{Osterwalder}K. Osterwalder and R. Schrader, \CMP{31}, 83 (1973);
              \CMP{42}, 281 (1975); \hfil\break
              K. Osterwalder and E. Seiler, \ANP{110}, 440 (1978).
\bibitem{footL} The poles corresponding to unstable states or resonances lie
              on the second Riemann sheet.
\bibitem{footM} $\hbar\ne0$ corresponds to non-zero temperature.
\bibitem{footN} For instance, there is a non-zero probability for all the
              molecules of air in a room to collect themselves in one corner,
	      but it would be silly to wait for such a thing to happen
	      spontaneously.
\bibitem{footO} Throughout this article, I stick to discussion of pure quantum
              states. This is with a view that mixed states can be made pure
	      by enlarging the Hilbert space of the system to include extra
	      degrees of freedom (e.g. those of the environment).
\bibitem{LSZ} H. Lehmann, K. Symanzik and W. Zimmermann, \NCX{1}, 1425 (1955).
\bibitem{footP} In fact, imposition of unitarity constraints on Euclidean
	      time results provides a way to quantify/bound the violations
	      of Bell's inequalities.
\bibitem{footQ} It is still possible to extract phase shifts from Euclidean
	      time results using indirect methods, e.g. finite volume shifts
	      in energy levels of a multi-particle system \cite{Luescher}.
	      Conversion of cross sections to phase shifts using the optical
	      theorem presupposes unitarity.
\bibitem{Luescher}M. L\"uscher, \CMP{105}, (1986) 153; \NPB{354}, 531 (1991).
\bibitem{Bell86}See for example: J. Bell, in {\it New Techniques and
              Ideas in Quantum Measurement Theory} (1986); reproduced
              as Chapter 21 in {\it Speakable and Unspeakable in
              Quantum Mechanics}, Ref.~\onlinecite{Bell64}.
\bibitem{footR} A density matrix with negative entries is perfectly all right
	      for quantum systems, though it would be anathema in statistical
	      physics.
\bibitem{footS} Though we are talking about the density matrix here, the
              situation is quite analogous to the evolution of the wavefunction               in Euclidean time: It is common knowledge that the ground state
	      wavefunction in quantum mechanics has no nodes. Euclidean time
	      evolution concentrates the wavefunction towards its ground
	      state component, cutting down the excited state contributions
	      responsible for nodes in the wavefunction.
\bibitem{Shamir} I thank Yigal Shamir for this remark.
\bibitem{BohmAharanov} D. Bohm and Y. Aharanov, \PRX{108}, 1070 (1957).
\bibitem{footT} The Wigner function for a spin-$\half$ particle (or any
	      other two state system) can be written as a $2\times2$ matrix.
	      The first index of this matrix can be the eigenstate index
	      (e.g. $\sigma_3$), and then the second index would be the
	      index for the operator generating transitions among the
	      eigenstates (e.g. $\sigma_1$).
\bibitem{FradkinPerelomov} See for instance: E. Fradkin, {\it Field Theories
              of Condensed Matter Systems} (Addison-Wesley, New York, 1991);\\
	      A. Perelomov, {\it Generalized Coherent States and Their
	      Applications} (Springer-Verlag, Berlin, 1986).
\bibitem{footU} $S^2 = SU(2)/U(1)$ forms the phase space for the dynamics of
              the ``classical spin''. The $U(1)$ bundle is responsible for
	      the geometric phase associated with the spherical area enclosed
	      during the spin's motion.
\bibitem{footV} When the generators are multiplied by a factor of $-i$, the
	      transformed set of elements don't even form a group. They can
	      be embedded in a bigger group such as $SL(2,C)$, however.
\bibitem{anomalies} Extensive discussion of anomalies in field theories can
	      be found in: S.B. Treiman, R. Jackiw, B. Zumino and E. Witten,
	      {\it Current Algebra and Anomalies} (World Scientific, Singapore,
	      1985).
\bibitem{Witten} E. Witten, \NPB{223}, 422 (1983).
\bibitem{Mermin98} N. D. Mermin, \AMJP{66}, 753 (1998), {\tt quant-ph/9801057}.
\bibitem{footW} With the standard feature of ${\cal CPT}$ invariance in
	      relativistic quantum field theories, $\CT-$violation and
	      ${\cal CP}-$violation are equivalent.
\bibitem{Berry} M.V. Berry, \PRSA{392}, 45 (1984).
\bibitem{ShapereWilczek} Many instances can be found in:
	      A. Shapere and F. Wilczek, {\it Geometric Phases in Physics}
	      (World Scientific, Singapore, 1989).
\bibitem{footX} According to Wigner's celebrated result \cite{Wignerbook},
	      symmetry transformations in quantum theory have to be either
	      unitary or antiunitary. The only known instance of antiunitary
	      transformation is time reversal, properties of which are
	      exploited in this article.
\bibitem{Wignerbook} E. Wigner, {\it Group Theory and its Applications
	      to the Quantum Mechanics of Atomic Spectra}
	      (Academic Press, New York, 1959).
\bibitem{Expt1} Q.A. Turchette, C.J. Hood, W. Lange, H. Mabuchi and
	      H.J. Kimble, \PRL{75}, 4710 (1995).
\bibitem{Expt2} C. Monroe, D.M. Meekhof, B.E. King, W.M. Itano and
	      D.J. Wineland, \PRL{75}, 4714 (1995).
\bibitem{Expt3}I.L. Chuang, L.M.K. Vandersypen, X. Zhou, D.W. Leung and
	      S.L. Lloyd, Nature {\bf 393}, 143 (1998).
\bibitem{bellineq93} A. Patel, ``{\it Another Look at Bell's Inequalities
	      and Quantum Mechanics}'', Contributed to the 16th International
	      Symposium on Lepton and Photon Interactions, Ithaca, August 1993,
              {\tt hep-th/9309077}.
\end{references}
\end{document}